\documentclass{appolb}
\usepackage{graphicx}
\usepackage{amsmath,amssymb}
\usepackage{epstopdf}
\usepackage{subfigure}

\begin{document}
\title{New Solvable Potentials with Bound State Spectrum
}
\author{K. Rajchel 
\address{Institute of Computer Science, Pedagogical University of Cracow, ul. Podchora\c{}\.{z}ych 2, PL- 30-084 Krak\'{o}w, Poland\\
krajchel@up.krakow.pl}
\\
}
\maketitle
\begin{abstract}
A new family of solvable potentials related to the Schroedinger-Riccati equation  has been investigated. This one-dimensional potential family depends on parameters and is restricted to the real interval. It is shown that this potential class, which is a rather general class of solvable potentials related to the hypergeometric functions, can be generalized to even wider classes of solvable potentials. As a consequence the nonlinear Schroedinger-type equation has been obtained.
\end{abstract}

\section{Introduction}
Though, we deal in the present paper with real equations the obtained results can be developed to the complex domain following the approaches \cite{Mi}.
Solvable problems of non-relativistic quantum mechanics have always attracted much attention \cite{Ju,Gd}.  The analytical methods to resolve the Schroedinger equation are very well-known \cite{BF}. A further remarkable development in solving the Schroedinger equation was the introduction of the concept of shape invariance \cite{Gd}. Many of the potentials related by supersymmetry \cite{Ju, Co, Ra} were found to have similar shapes (i.e. to depend on the coordinate in similar way), only the parameters appearing in them were different. Although the number of potentials satysfying the shape invariance condition is limited, it turned out that the energy spectrum and the wavefunctions can be determined by elementary calculations in this case.

\section{Schroedinger equation}
In the present paper we consider the Schroedinger equation in one dimension, setting $\hbar=2m=1$:

\begin{equation}
\-{d^2 \over dx^2}\psi_n(x)+ (E_n-V(x)) \psi_n(x)=0
\label{e1}
\end{equation}
then the function

\begin{equation}
W_n(x)=-{\psi_n'(x) \over \psi_n(x)},
\label{e2}
\end{equation}
where prime denotes differentation with respect to $x$, satisfies the corresponding Riccati equation

\begin{equation}
W'_n(x)-W_n^2(x)=E_n-V(x).
\label{e3}
\end{equation}
Assuming that the function $W_0(x)$ has a zero inside interval $I$ and

\begin{equation}
W_0'(x)>0 \quad \forall x \in I \subset \mathbb{R},
\label{e4}
\end{equation}
which is associated with normalization of the basic function $\psi_0$, we get
\begin{equation}
W_0'(x)=F(W_0),
\label{e5}
\end{equation}
where $F$ is an arbitrary function satisfying eq.\eqref{e4}.
The last equation is obtained from reversibility of the function $W_0(x)$ on interval $I$. 
Taking eq.\eqref{e5} into account and comparing it with eq.\eqref{e3} we get the following result:

\begin{equation}
W_0'(x)=W_0^2+f(W_0),
\label{e6}
\end{equation}
where
\begin{equation}
E_0-V(x)=f(W_0).
\label{e7}
\end{equation}
Now we can express the potential $V(x)$ in terms of $W_0$ and we can use eq.\eqref{e6} to generate potentials by choosing $f(W_0)$.

A simplest and most obvious choice seems to be a second order polynomial

\begin{equation}
W_0'(x)=AW_0^2+BW_0+C,
\label{e8}
\end{equation}
where $A, B, C$ are parameters.
This differential equation is a first-order one and it can be solved in a straightforward way.
The solution of the eq.\eqref{e8} has the form

\begin{equation}
W_0(x)=-{B \over 2A}+{\sqrt{-B^2+4AC} \over 2A}\tan\left({1\over 2}\sqrt{-B^2+4AC}(x-x_0)\right),
\label{e9}
\end{equation}
where
\begin{equation}
x_0-{\pi \over \sqrt{-B^2+4AC}}\leq x \leq x_0+{\pi \over \sqrt{-B^2+4AC}}.
\label{e10}
\end{equation}
Thus
\begin{equation}
\psi_0(x)=e^{B(x-x_0)\over{2A}} (\cos({1\over 2}\sqrt{-B^2+4AC}(x-x_0))^{1\over A}
\label{e11a}
\end{equation}
is the unnormalized form of the ground state.

\section{ Cascade of equations}

In order to have the same potential function in the Riccati equation for $n=1$ we introduce expression for $W_1$ :
\begin{equation}
W_1=W_0-{a_1\over b_1W_0-c_1}.
\label{e11}
\end{equation}
Now, let us turn our attention to the explicit determination of the coefficients $a_1, b_1, c_1$ in terms of the $A, B, C$. From the eq.\eqref{e3} and eq.\eqref{e8} we get
\begin{equation}
a_1=(A+2)C-{(A+2)^2B^2\over 4(A+1)^2},
\label{e12}
\end{equation}
\begin{equation}
b_1=A+2,
\label{e13}
\end{equation}
\begin{equation}
c_1=-{(A+2)B \over 2(A+1)}.
\label{e14}
\end{equation}
Strightforward calculations lead us to the form of the first excited state wavefunction
\begin{equation}
\psi_1(x)=e^{\alpha_{01} x} (\cos\theta(x-x_0))^{\gamma_{01}} (\alpha_{1} \cos\theta(x-x_0)+\beta_{1} \sin\theta(x-x_0))^{\gamma_{1}},
\label{e16a}
\end{equation}
where  all coefficients, denoted in the Greek letters, depend on  the parameters $A, B, C$. Although this relationship is rather complex but, by use of the equations \eqref{e2}, \eqref{e9} and \eqref{e11}, easy to achieve.

Considerations presented above can be generalized if we take the explicit form of the function $W_n$ in terms of $W_0$ :
\begin{equation}
W_n=W_0-\cfrac{a_n}{b_nW_0-\cfrac{c_n}{b_{n-1}W_0-\cdots}}.
\label{e15}
\end{equation}
This function preserves the expression of equation
\begin{equation}
W'_n(x)-W_n^2(x)= W'_0(x)-W_0^2(x)+ E_n-E_0
\label{e16b}
\end{equation}
for suitable values of coefficients which are involved in a system of non-linear equations (too complicated to be presented here). We should select  the appropriate values of  $A, B, C$ parameters to simplify calculations. It will be done in the next chapter.

 The equations \eqref{e2}, \eqref{e3} and \eqref{e8} enable us to obtain every wavefunctions $\psi_n$ and energies $E_n$. It is easy to prove \cite{Ra} that the function $W_0$ fulfil the shape invariance condition, so this potential family, resulting from the eq.\eqref{e8}, is an example for shape-invariant solvable potentials.
Using eq.\eqref{e15} the wavefunctions can be written, without normalization, as
\begin{equation}
\psi_n(x)=e^{\alpha_{0n} x} (\cos\theta(x-x_0))^{\gamma_{0n}}\prod_{i=1}^{n} (\alpha_{i} \cos\theta(x-x_0)+\beta_{i} \sin\theta(x-x_0))^{\gamma_{i}},
\label{e16}
\end{equation}
where, as in the previous case, all coefficients written in Greek depend on  the  $A, B, C$ parameters.

\section{The classic potentials}

Equation \eqref{e8} offers a convenient way to link this simple method with the well-known solutions of the Schroedinger equation. 
For instance, choosing $B=0, C=1$ we get the following results:
\begin{equation}
W_0'=AW_0^2+1
\label{e17}
\end{equation}
and the first three wavefunctions
\begin{equation}
\psi_0(x)=(\cos{(\sqrt{A}x)})^{1\over A},
\label{e18}
\end{equation}
\begin{equation}
\psi_1(x)={(\cos{(\sqrt{A}x)})^{1\over A}\sin{(\sqrt{A}x)}\over \sqrt{A}},
\label{e19}
\end{equation}
\begin{equation}
\psi_2(x)={(\cos{\sqrt{A}x)})^{1\over A}(-1+(1+A)\cos{(2\sqrt{A}x)})\over A},
\label{e20}
\end{equation}
which are orthoghonal on domain $I$ (eq.\eqref{e10}), tend to very well-known solutions of the quantum oscillator $\psi_n(x)\to H_n(x)e^{-{x^2\over 2}}$ for $A\to0$.
It can be seen from the figures below that the wavefunctions have the same characteristic shapes but they differ in domains.
\pagebreak
\begin{figure}[h]
\begin{center}
\subfigure[ The first three wavefunctions  for $A>0 \quad (A=0.9)$.]{
\resizebox*{5cm}{!}{\includegraphics{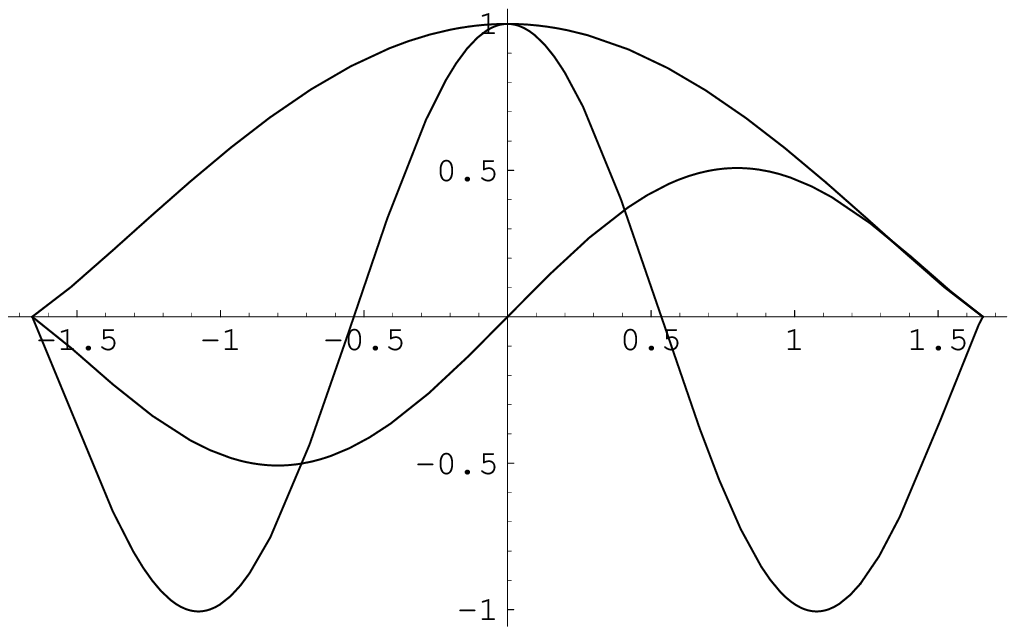}}}\hspace{5pt}
\subfigure[The first three wavefunctions of the quantum harmonic oscillator $(A\to 0)$.]{
\resizebox*{5cm}{!}{\includegraphics{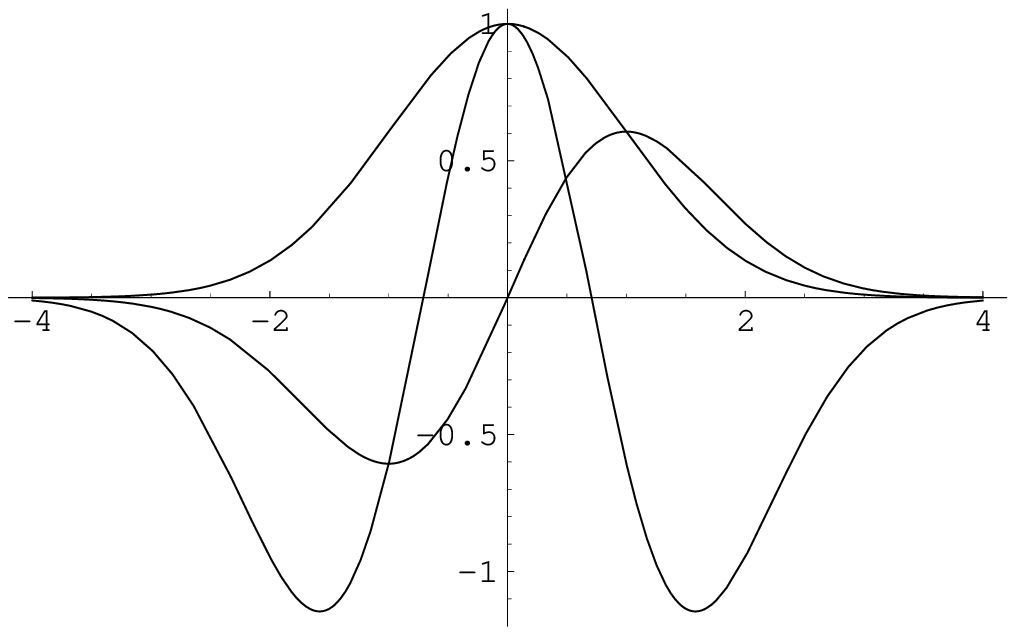}}}
\caption{Example of the wavefunctions with different values of the parameter $A$, where $x$ is on the horizontal axis and $\psi_n(x)$ on the vertical one.}
\label{figure1}
\end{center}
\end{figure}

Basing on the procedure described above we are able to get solutions of the Schroedinger equation with the radial Coulomb potential (angular momentum is equal to zero). In this case
\begin{equation}
W_0'=AW_0^2-BW_0+{B^2\over 4}
\label{e21}
\end{equation}
whose basic solution is
\begin{equation}
\psi_0(x)={(\sin({{1\over 2}\sqrt{A-1}}Bx))^{1\over A}\over \sqrt{A-1}}e^{-{B\over 2A}x} 
\label{e22}
\end{equation}
which tends to the radial part of the ground state eigenfunction of the Schroedinger equation for  one-electron atom, $\psi_0\to{1\over 2 }Bxe^{-{1\over 2}Bx}$ for $A\to1$. With help of the eq.\eqref{e15} we are able to get the wavefunctions for the excited states.

Another example of the eq.\eqref{e8} which lead us to the very well-known solution is
\begin{equation}
W_0'=-AW_0^2-W_0+C,
\label{e23}
\end{equation}
where parameters $A>0$,  and $C>0$.
Thus
\begin{equation}
W_0(x)=-{1\over 2A}+{\sqrt{1+4AC}\over2A}\coth\left({1\over 2}\sqrt{1+4AC}(x-x_0)\right),
\label{e24}
\end{equation}
where the integration constant
\begin{equation}
x_0={1\over \sqrt{1+4AC}}(\ln{A}+\imath \pi)
\label{e25}
\end{equation}
is the complex number.
In this case we have
\begin{equation}
W_0(x)=-{1\over 2A}+{\sqrt{1+4AC}\over2A}\tanh\left({1\over 2}(\sqrt{1+4AC}x-\ln{A})\right)
\label{e26}
\end{equation}
and
\begin{equation}
\psi_0(x)=e^{x\over 2A}\cosh\left({1\over 2}(\sqrt{1+4AC}x-\ln{A})\right)^{-{1\over A}}
\label{e27}
\end{equation}
is the basic wavefunction without normalization constant.
From eq.\eqref{e26} we have $W_0=C-e^{-x}$ for $A\to0$ what is the standard expression for the Morse potential \cite{Co}.

All potentials resulting from eq.\eqref{e8} have a trigonometric form. It means that they are expressed in terms of the tangent function. If we wish to obtain potentials interesting from the physical point of view, like the Coulomb potential, or the Morse potential, we should follow the procedure outlined above or choose the proper, inital value of the parameters in eq.\eqref{e8}. It should be emphasized that every solution of the Schroedinger equation related to the orthogonal polynomials can be obtained by this method.

\section{The new Hamiltonian}

The results can be generalized to the form for which this method works \cite{RS}. If we take eq.\eqref{e8}, in the form
\begin{equation}
W_0'=AW_0^2+{P_{l+1}(W_0)\over Q_{l}(W_0)}=R_{l+2,l}(W_0),
\label{e28}
\end{equation}
where $P_{l+1}(W_0)$ is a polynomial in $W_0$ with degree no greater than $l+1$, $Q_{l}(W_0)$ is a polynomial in $W_0$ with degree equal to $l$ and $R_{l+2,l}(W_0)$ is a rational function such that both the numerator and the denominator are polynomials with degree $l+2$ and $l$ respectively.
Substituting
\begin{equation}
W_0(x)=R_{l+1,l}(\tan(\phi \cdot(x-x_0)))
\label{e29}
\end{equation}
in eq.\eqref{e28} and adjusting the indices of the sums to get the same powers\\ of $\tan(\phi \cdot(x-x_0))$, we get the explicit form of $W_0$. It should be emphasized that the condition of eq.\eqref{e4} must be satisfied. The procedure outlined above can be applied to the function
\begin{equation}
W_n(x)=R_{l+n+1,l+n}(\tan(\phi \cdot(x-x_0)))
\label{e30}
\end{equation}
and thus the excited state wavefunctions $\psi_n(x)$ can be obtained.

Let us consider the simple equation, being the example of generalized eq.\eqref{e28}
\begin{equation}
W_0'=W_0^2+{3W_0-1\over W_0-3}={(W_0-1)^3\over W_0-3},
\label{e31}
\end{equation}
where all coefficients has been chosen to simplify calculations.\\
Hence
\begin{equation}
W_0(x)={2\sqrt{x+{1\over 4}}-3\over 2\sqrt{x+{1\over 4}}-1},
\label{e32}\end{equation}
 which gives the correspondig eigenvalue $E_0=-1$ and the completly new potential is discovered:
\begin{equation}
V(x)=-{2\over \sqrt{x+{1\over 4}}}\quad\textrm{for }x\geq 0.
\label{e33}
\end{equation}
Thus the ground state function, without normalization constant, has the form
\begin{equation}
\psi_0(x)=e^{-x+ 2\sqrt{x+{1\over 4}}}\left( 2\sqrt{x+{1\over 4}}-1\right).
\label{e34}
\end{equation}
Substituting
\begin{equation}
W_1={P_2(W_0)\over Q_1(W_0)}
\label{e35}
\end{equation}
in eq.\eqref{e3} and taking into account eq.\eqref{e33}, we obtain the unnormalized wavefunction
\begin{equation}
\psi_1(x)=e^{-0.79x+ 2.52\sqrt{x+{1\over 4}}}\left( 2\sqrt{x+{1\over 4}}-1\right)\left( 2\sqrt{x+{1\over 4}}-3.74\right),
\label{e36}
\end{equation}
where all decimal numbers are approximated and $E_1\approx-0.63$. It is easy to show that the latest potential does not fulfil the shape invariance condition \cite{Ra}, so this new potential family is an example for non-shape-invariant solvable potentials.

Let us now discuss the question of the explicit form of the Schroedinger equation. Treating the eq.\eqref{e28} not as a condition but rather as the transformed Schroedinger equation and substituting eq.\eqref{e2} (for $n=0$) to eq.\eqref{e28}, we get
\begin{equation}
-\psi_0''(x;\alpha)\psi_0(x;\alpha)-\alpha(\psi_0'(x;\alpha))^2=\left[ E_0-V\left(-{\psi_0'(x;\alpha)\over \psi_0(x;\alpha)}\right)\right]\psi_0^2(x;\alpha),
\label{e37}
\end{equation}
where the parameter $\alpha$ is usually related to the parameter $A$ in eq.\eqref{e28} and 
the potential $V$ has the form
\begin{equation}
V\left(-{\psi_0'(x;\alpha)\over \psi_0(x;\alpha)}\right)=R_{l+2,l}\left(-{\psi_0'(x;\alpha)\over \psi_0(x;\alpha)}\right).
\label{e38}
\end{equation}
As we see the eq.\eqref{e37} is nonlinear differential equation. Taking into account the previous considerations regarding the quantum oscillator, the Coulomb potential and the Morse potential we get the following form of the eq.\eqref{e37} in the $\alpha \to0$ limit:
\begin{equation}
-\psi_0''(x;0)=\left[ E_0-V\left(-{\psi_0'(x;0)\over \psi_0(x;0)}\right)\right]\psi_0(x;0)
\label{e39}
\end{equation}
what is the familiar form of the Schroedinger equation, and where nonlinearity is hidden in the form of the potential function.
\section{Conclusions}

The new method of obtaining solvable potentials has been reviewed in this paper. The main role in this method plays the Riccati equation which is a result of the transformed, one-dimensional, stationary Schroedinger equation. It allows us to emphasize the importance of function $W_0$ known in a literature as a "superpotential" \cite{Ju,Co}. By use of its features we can show that the potential is not an arbitrary function of $x$ but rather its form depends on the function $W_0$. As a consequence, we can find not only very well-known solutions of the Schroedinger equation but also a new class of the solvable potentials. These considerations may help us to identify new clasess of the solvable potentials and may serve as an aid for further investigations concerning the relationship between solvability of the Schroedinger equation and the form of the potential.

\end{document}